# R&D ON BEAM INJECTION AND BUNCHING SCHEMES IN THE FERMILAB BOOSTER *

C. M. Bhat#
Fermilab, Batavia, IL 60510, USA

*Abstract*

Fermilab is committed to upgrade its accelerator complex to support HEP experiments at the intensity frontier. The ongoing Proton Improvement Plan (PIP) enables us to reach 700 kW beam power on the NuMI neutrino targets. By the end of the next decade, the current 400 MeV normal conducting LINAC will be replaced by an 800 MeV superconducting LINAC (PIP-II) with an increased beam power >50% of the PIP design goal. Both in PIP and PIP-II era, the existing Booster is going to play a very significant role, at least for next two decades. In the meanwhile, we have recently developed an innovative beam injection and bunching scheme for the Booster called "early injection scheme" that continues to use the existing 400 MeV LINAC and implemented into operation. This scheme has the potential to increase the Booster beam intensity by >40% from the PIP design goal. Some benefits from the scheme have already been seen. In this paper, I will describe the basic principle of the scheme, results from recent beam experiments, our experience with the new scheme in operation, current status, issues and future plans. This scheme fits well with the current and future intensity upgrade programs at Fermilab.

## INTRODUCTION

Nearly one and a half decades ago, Fermilab started focusing on upgrades to its accelerator complex towards the intensity frontier that would substantially increase the average beam power delivered to the fixed target HEP experiments (as well as support then ongoing ppbar collider program) thereby transforming the facility into a world class accelerator based neutrino facility.

Currently, the chain of accelerators in the complex consists of an RFQ, 400 MeV normal conducting RF LINAC, 0.4-8 GeV rapid cycling Booster, 8 GeV permanent magnet Recycler Ring and 8-120 GeV (or 150 GeV) Main Injector. The last three machines in this chain are synchrotrons. The primary goal of the upgrades was delivering 700 kW of beam power at 120 GeV on the NuMI/NOvA target (a high energy neutrino experiment), and simultaneously provide proton beams to the low energy neutrino and fixed target experiments.

In 2010, after two and a half decades of successful operation of the Tevatron ppbar collider, the energy frontier HEP programs moved to the LHC at CERN. Since then many new developments have taken place at Fermilab. The Recycler, originally used as the primary anti-proton storage ring during the Tevatron collider era, has been upgraded to a high intensity proton storage ring that can be used as an injector to the Main Injector. This increased the Main Injector duty factor by nearly 30%. Though the Fermilab Booster is one of the oldest rapid cycling proton synchrotron in the world [1, 2] that cycles at 15 Hz and is in operation since 1971, until 2002 it delivered the beam on average at a rate of ≈1 Hz or less with a maximum beam intensity of ~3.5E12 p per Booster cycle (ppBc). During 2002-15 the beam delivery rate from the Booster has been increased to about seven cycles per sec as MiniBooNE and MINOS came online. The PIP was established around 2010 [3] to support the newly proposed NOvA, g-2, Mu2e, and short-baseline neutrino experiments which demanded doubling the Booster beam repetition rate from 7.5 Hz to 15 Hz with about 4.6E12 ppBc. The foreseen Proton Improvement Plan-II [4, 5] supports the long-term physics research programs by providing MW type beam power to LBNE while sending beam to the on-going HEP experiments and forms a platform for the future of the Fermilab. The main components of the PIP-II are a new 800 MeV superconducting LINAC as an injector to the Booster and increase the Booster beam delivery repetition rate to 20 Hz with about 6.7E12 ppBc. In any case, the Booster is going to play a very important role at least for the next two decades and will remain the workhorse in the Fermilab accelerator complex.

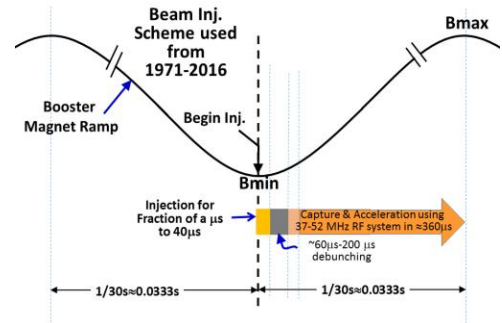

Figure 1: Schematic of the beam injection scheme in operation for the past forty five years of the Booster.

Booster uses sinusoidal magnetic ramp for beam acceleration. Its cycle rate is locked to 60 Hz ComEd power distribution system. The Booster has a circumference of 473.8 m with 96 combined function magnets distributed on a FOFDOOD (DOODFOF) 24 symmetric lattice period with independently controllable power supplies to its correctors to control its transverse dynamics. The fundamental accelerating RF system operates with a harmonic number h=84 and sweeps its frequency in the range of 37.8 to 52.8 MHz in 1/30$^{th}$ of a second during the beam acceleration. The beam from the LINAC arrives at the Booster with a 200 MHz bunch structure. At the beginning the Booster was operated with single turn proton injection [6] and since 1978 Booster has adopted $H^-$ multi-turn charge exchange injection technique [7].

_________________________________________
* Work supported by Fermi Research Alliance, LLC under Contract No. De-AC02-07CH11359 with the United States Department of Energy
# cbhat@fnal.gov

Until the end of 2015 the Booster received the beam at the minimum of its magnetic field, $B_{min}$ as shown in Fig. 1. Irrespective of the length of the LINAC beam pulse (< 40 µsec) the injected beam was allowed to debunch for a period of about 60-200 µsec and captured subsequently. Since the magnetic field was continuously increasing the beam was captured as quickly as possible with considerably large RF buckets. In addition to this, the fluctuation of the ComEd power line frequency which is of the order of 100 mHz out of 60 Hz introduced both time jitter (~50 µsec) and amplitude jitter in $B_{min}$ (leading to ~0.5 MeV fluctuation). Also, as shown in Fig. 1, the beam capture and acceleration found to partly overlap during this part of the cycle. A combination of all these effects led to undesirable beam filamentation in the RF buckets leading to longitudinal emittance dilution, decreased beam capture efficiency and possibly transverse emittance growth at injection which might mimic space charge related issues. This also puts severe limits on achievable beam intensity. Over the years many improvements have been added to make the beam operation more efficient. Yet, the best efficiency observed so far was <95% with the scheme shown in Fig. 1 and a substantial longitudinal emittance dilution.

In 2014, we proposed a new injection scheme [8] called *Early Beam Injection scheme* (EIS), which fits well between PIP and PIP-II eras and has high potential to increase the beam power significantly. This scheme involves beam injection on the deceleration part of the magnet ramp in the Booster. At the end of 2015 we have implemented the new scheme in operation. Here, we explain briefly the general principle of the scheme, the results from beam dynamics simulations and beam studies, the current status of the scheme in operation, and future prospects.

## EARLY INJECTION SCHEME

A schematic view of the newly proposed injection scheme is shown in Fig. 2. The basic idea of this scheme is to inject and capture the beam on a *pseudo front porch* created by imposing $dP/dt = 0$ in a changing magnetic field. Conventional wisdom was that this is not possible unless there is a front porch with a constant magnetic field. We noticed that around the minimum (and maximum) of an ideal sinusoidal dipole magnetic ramp the field changes slowly and symmetrically. Therefore, one can start injecting the beam relatively earlier than $B_{min}$. In a decreasing magnetic field the injected beam with a fixed energy starts moving towards the outside of the ring (injection energy is below the transition energy of the Booster). In the case studied here, the beam injection is carried out at ≈ 200 µsec prior to $B_{min}$. For the Booster parameters (shown in Table 1) the maximum radial displacement of the beam centroid is ≈1 mm due to change in magnetic field, which is << 57.2 mm, the limiting physical aperture (the diameter of the RF cavity iris opening). The injection process itself takes as much time as the length of a LINAC pulse. Immediately after the completion of the injection, the Booster RF system is turned-on at a frequency matched to the beam revolution frequency. Debunching of the beam prior to the start of beam capture is eliminated. $dP/dt = 0$ is imposed by keeping radial feed-back turned off till the end of capture. Changing $B$ field at a constant momentum still introduces varying revolution frequency in accordance with $\Delta B / B = \gamma_T^2 \Delta f / f$, where $\gamma_T$ =5.478 is transition gamma for the Booster. The corresponding change in the RF frequency is ≈15.1 kHz. Thus, on the deceleration ramp the required RF frequency decreases initially and reaches its minimum at $B_{min}$ and increases symmetrically. This RF frequency variation should be taken in to account during the beam capture though the beam radially swings outside and inside. The beam is captured by increasing the RF voltage from about 20 kV to 400 kV in about 240 µsec. At the same time the beam synchrotron frequency changes from about 6 kHz to 27 kHz. In an ideal case, one demands much longer capture time. Since, the magnetic field is changing continuously and also the beam is moving radially during the capture, the time required to capture the beam cannot be increased much further.

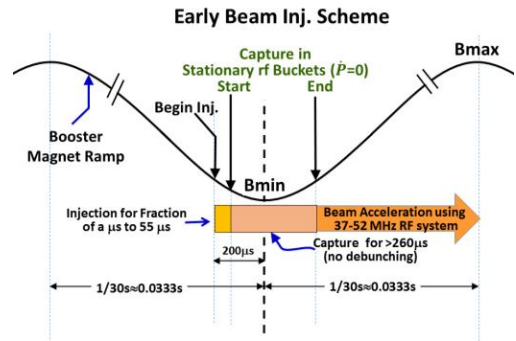

Figure 2: A schematic view of the new injection scheme in the Fermilab Booster.

The energy spread, Δ*E* (full), of the incoming multi-turn beam is about 1.25 ±0.20 MeV [9]. On the other hand, the bucket height from the residual RF voltage of nearly 20 kV is 0.9 MeV which is smaller than the energy spread of the injected beam. Hence, though the bunching starts immediately after the beam arrives into the Booster, the emittance dilution due to non-zero RF voltage is very small. (If the initial bucket height is comparable or larger than the energy spread of the incoming beam then one expects noticeable emittance dilution at capture.) By the completion of the capture the beam energy spread goes up to 3.6 MeV and the beam bunches will be on the increasing part of the magnetic field ramp. This beam energy spread is still smaller than the Booster energy acceptance at injection which is ~5.4 MeV [10], hence, we do not anticipate any beam losses during the beam capture. RF feedback is turned-on for beam acceleration on the fully bunched beam.

Table 1: Booster parameters used in the simulations

| Parameters | |
|---|---|
| Booster circumference ($2\pi R$) [m] | 473.8 |
| Injection KE [MeV] | 400 |
| Extraction KE [MeV] | 8000 |
| Cycle Time[sec] | 1/15 |
| Beam injection w.r.t. $\dot{B} = 0$ [μsec] | -200 |
| Harmonic Number | 84 |
| Transition Gamma $\gamma_T$ | 5.478 |
| RF Frequency [MHz] | 37.8-52.8 |
| Beam Structure at Injection | 201MHz |
| LINAC Pulse length [μsec] | 36-50 |
| Number of Booster Turns | 16-22 |
| $\Delta E$ at Injection [MeV] | 1.25 [9] |
| $\varepsilon_L$ at injection/84 bunches [eV sec] | 2.77 |
| $\varepsilon_L$/bunch [eV sec] | 0.033 |
| Bunch Intensity [protons/bunch] | 2E10-15E10 |
| Beam transverse radius [cm] | 1.2 |
| Beam pipe (RF) radius [cm] | 2.86 |

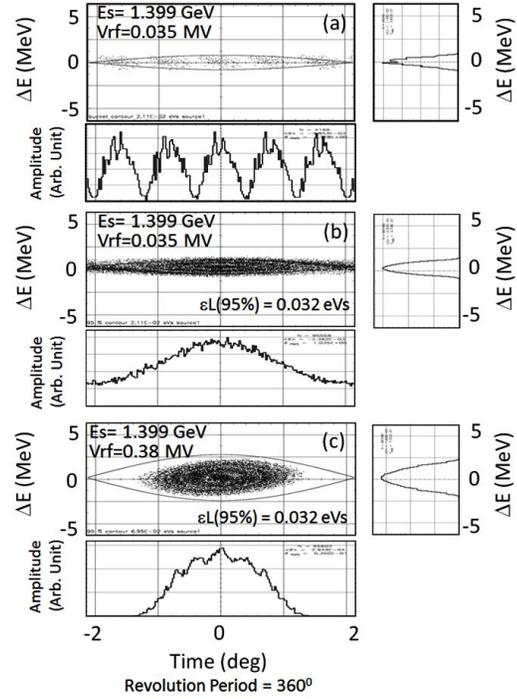

Figure 3: Simulated phase space distributions for the first 260 μsec, a) LINAC beam on the first turn in the Booster, b) at completion of 22 Booster turn injection, c) completion of beam capture in 37 MHz Booster RF bucket. The line-charge distribution and the predicted energy distributions are also shown on the right hand side.

We have demonstrated the feasibility of the EIS in the Booster using 2D- particle tracking simulation code ESME [11] including the longitudinal space charge effects. Table 1 lists the machine and the beam parameters used in the simulations. Figures 3-6 display the results from simulations for 9E10 p/Booster bunch which is about 70% larger than the PIP design intensity.

Simulations showed that there is a small longitudinal emittance dilution during the beam capture and that emittance is preserved till the transition energy. The dilution mainly comes from the non-zero RF voltage at injection. The transition crossing adds further emittance dilution; the full emittance increases by 70% from 0.048 eVs to 0.083 eVs. Majority of this arises from RF bucket mismatch as shown in Fig. 5. Interestingly, the simulations showed that the 95% emittance did not change much.

The Recycler Ring uses a multi-batch slip stacking technique [12] to increase the proton flux and it demands full beam energy spread from the Booster to be <13 MeV. The beam energy spread at the end of acceleration is about 20 MeV as shown in Fig. 6(a). To reduce the energy spread to an acceptable value by the Recycler Ring we adopt snap bunch rotation rather than currently used quadrupole RF voltage modulation [13]. The results from the simulations on snap bunch rotation are shown in Fig. 6(b). One can minimize any observed distortion in the rotated phase space distribution of the beam particles by adding 16% of 2$^{nd}$ harmonic RF component to the fundamental 53 MHz waveform to linearize the effective waveform during the bunch rotation.

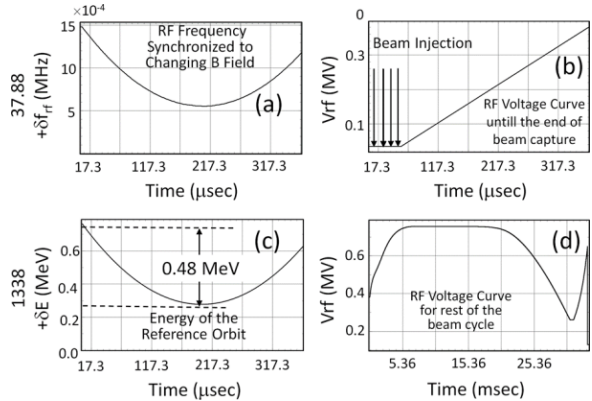

Figure 4: (a) Predicted variation of the RF frequency, (b) required RF voltage curve and (c) energy of the reference orbit, which represents the radial motion of the beam particles in the dipole field for the first 350 μsec. (d) predicted RF voltage curve for the entire acceleration cycle.

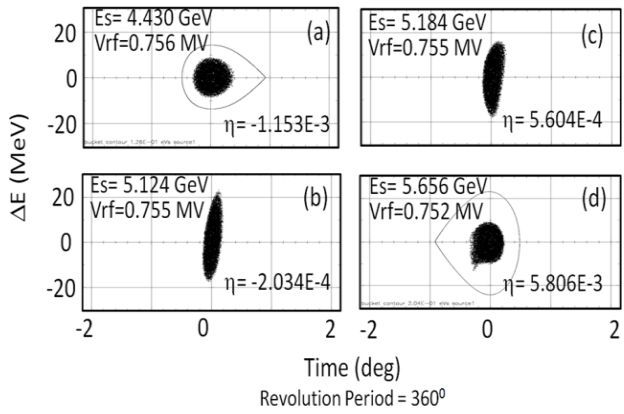

Figure 5: Simulations for the transition crossing. Distributions (a) before transition crossing, (b) and (c) very close to transition energy, (d) away from transition energy. We can see bucket mismatch.

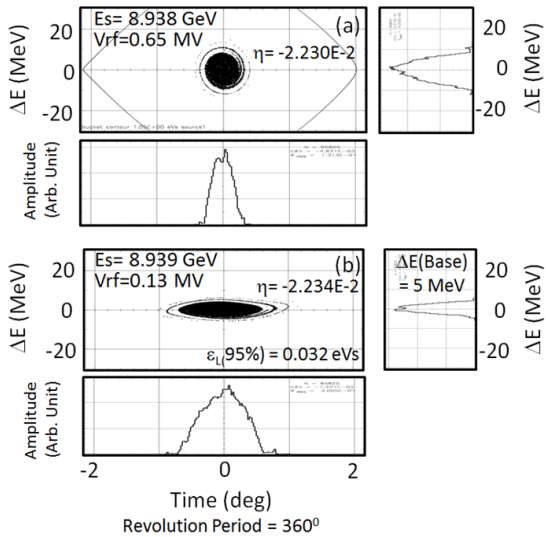

Figure 6: Simulation for snap bunch rotation: (a) before rotation commences and (b) at the end of rotation.

## EXPERIMENTS

Proof of principle experiments have been carried out along the lines of simulations on the EIS. The top picture of Fig. 7(a) displays the measured data for the first 1 msec for beam injection, capture and the early part of the acceleration for ≈5.6E12 ppBc. Zero crossing of the Bdot curve (same as the *Bmin*) occurring at ≈ 200 μsec after the beam injection is also shown for clarity. We also show an approximate timing of the acceleration turn on in this figure. The beam transmission efficiency for the first 1 msec is found to be about 97%.

The data on various beam intensities under similar conditions but, for the entire cycle are shown in Fig. 7 (b). The observed sudden step loss at the beginning of each case is due to a notch created soon after the beam capture. (This notch keeps rise time of extraction kicker cleared from any beam.) This apparent decrease in efficiency is

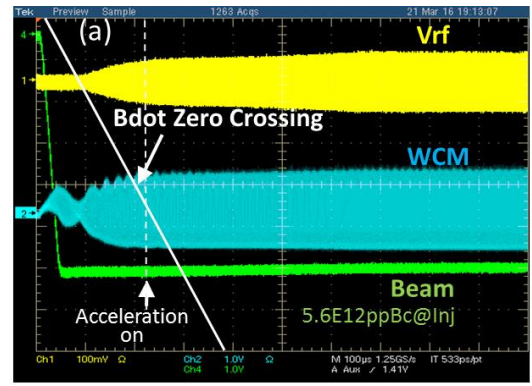

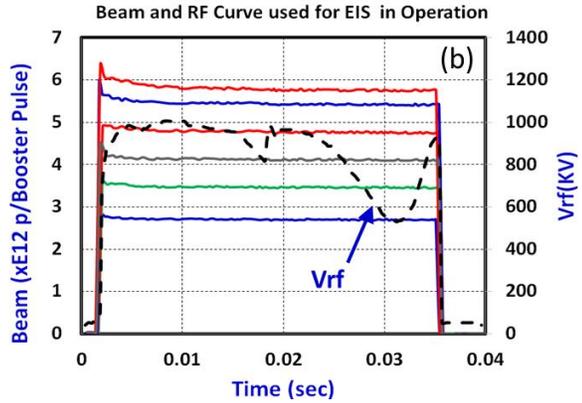

Figure 7: (a) Measurement data on first 1 msec at injection. (b) The beam through the acceleration cycle for different beam intensities. RF voltage curve is also shown here (dashed curve).

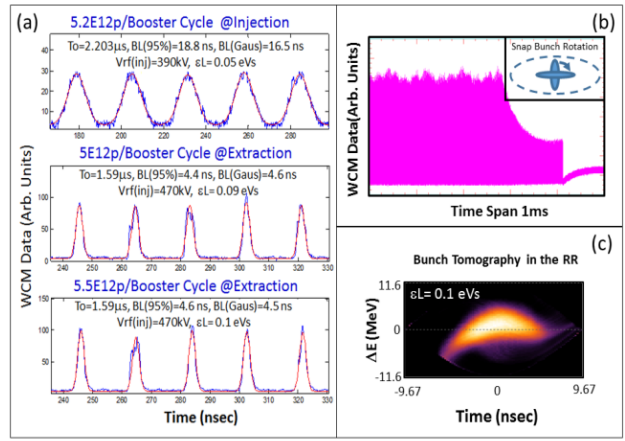

Figure 8: Measured wall current monitor (WCM) data on (a) five bunches at the end of beam capture and extraction for two intensities, (b) during the snap bunch rotation (schematic of bunch rotation shown in inset) and (c) the measured bunch tomography in the Recycler Ring for the beam coming from the Booster after bunch rotation.

~3.5%. These data show an average acceleration efficiency ~ 95% even for higher beam intensities.

Figure 8 (a) displays the bunch profiles and the measured longitudinal emittances at the end of the capture and also just before the extraction. The top two bunch traces are obtained on the same injected beam while the bottom trace is for a case with higher beam intensity. Soon after the end of the beam capture the 95% emittance was found to be ≈ 0.05 (±15%) eVs which is about 50% larger than that expected from the simulations. About 400 μs before the extraction the 95% longitudinal emittance ≈ 0.1 (±15%) eVs per bunch for both intensities. Figure 8(b) shows the wall current monitor data taken during the snap bunch rotation on a beam with 5.5E12 ppBc. The decreasing amplitude of the wall current monitor signal is the result of increasing bunch length. We do not see any particle falling out of the buckets during this time. A tomoscope reconstruction of the phase space distribution of the beam particles transferred from the Booster to the Recycler 53 MHz RF bucket is shown in Fig. 8(c). The measured 95% emittance and the 1σ energy spread are about 0.1 eVs and 2.83 MeV, respectively. This emittance is consistent with that measured in the Booster at extraction. This energy spread is about 10% less than that generally obtained in the current operation (notice that operationally we use about 20% less beam particles per bunch than the one illustrated here).

As of December of 2015, we have replaced the old injection scheme in the Booster with the early injection scheme and gaining operational experience. Even with partial implementation of the EIS in operation we have seen a few advantages, *e.g.*, i) the beam longitudinal emittance delivered from the Booster to the Recycler or the Main Injector has improved by >10%, ii) the average RF power per Booster cycle has also gone down by 10-15% as compared with the old scheme and iii) we were able to send higher intensity beam to the down-stream facilities. Since the implementation of the EIS a number of other improvements were also added as part of the PIP plans. The Booster beam delivery rep-rate has been increased from 7.5 to 15 Hz. We were able to deliver up to 701 kW beam power on the NuMI/NOvA target, recently.

## ISSUES, MITIGATION AND FUTURE PROSPECTS

There are a number of issues yet to be solved to take full advantages of the EIS in operation. Some of these are: 1) As mentioned earlier, the jitter in the $B_{min}$ relative to the beam injection clock event is quite large. This jitter arising from ComEd power line frequency is random and introduces large uncertainty during the start of adiabatic beam capture, there by emittance dilution in the early part of the cycle. Furthermore, this jitter also introduces uncertainty during transition crossing leading to RF phase mismatch and large quadrupole oscillation after transition crossing [14]. 2) The RF frequency does not follow the Booster dipole magnetic field ramp. (3) A better RF voltage regulation is needed at injection. Any unwanted imbalances in the RF voltage vectors introduces emittance dilution. As a consequence of these issues, we see longitudinal emittance dilution of about 50% during the beam capture. Transition crossing introduces another factor of two emittance increase. Currently R&D is in progress to mitigate every one of the above mentioned problems.

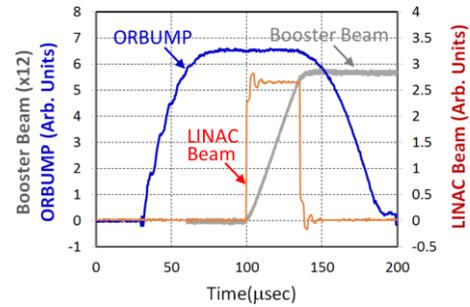

Figure 9: Current ORBUMP in the Booster at injection, LINAC pulse and injected and circulating beam with 5.8E12ppBc.

Table 2 summarizes the PIP and PIP-II performance goals. It also shows our expectation with full implementation of the EIS in operation after addressing the issues described earlier (which is important even for the PIP-II success). We find that the EIS fits well between PIP and PIP-II plans. With 6.4E12 ppBc at injection one can achieve ≈950 kW beam power on the NOvA target.

Table 2: PIP, PII-II parameters and expected from EIS.

| Parameters | PIP (**EIS***) | PIP-II |
|---|---|---|
| Inj. Energy (K.E.) | 0.4 (**0.4**)GeV | 0.8 GeV |
| Energy at Exit (K.E.) | 8 (**8**) GeV | 8 GeV |
| Booster Rep-Rate | 15 (**15**)Hz | 20 Hz |
| LINAC Pulse Length | ≈30 (**45**) μsec | 600 μsec |
| Intensity@Inj (ppBc) | 4.52(**6.4**) E12 | 6.63E12 |
| Inj. to Exit Efficiency | 95% (**>97%**) | 97% |
| Beam Power@Exit | 94 (**≈135**) kW | 184kW |
| Power@NOvA Target | 700(**≈950**)kW | 1.2 MW |

*PIP with EIS.

The EIS in principle can accommodate 60% longer $H^-$ pulses than the currently being used. The current LINAC can provide stable beam of about 50 μsec long pulses at 25 mA [15]. As shown in Fig. 9, the injection ORBUMP is wide enough to allow such a long beam pulse into the Booster. Thus, by using a longer LINAC beam pulses one should be able to increase the beam intensity beyond that mentioned above. We also do not anticipate any significant transverse emittance dilution due to multiple passage of the circulating beam through the stripping carbon foil [16]. In conclusion, EIS in the Booster has a high potential for increasing the beam intensity output by >40% than the PIP design with no/minimum beam loss.

I would like to thank W. Pellico, C. Drennan, K. Triplett, S. Chaurize, K. Seiya, B. Hendricks, T. Sullivan, F. Garcia, and A.Waller for many useful discussions their help in the beam studies. Special thanks are due to Fermilab Accelerator Division MCR crew.